# Observational Signatures of Self-Destructive Civilisations

**Authors:** Adam Stevens[1,2], Duncan Forgan[3] and Jack O'Malley James[3,4]

1: Department of Physical Sciences, The Open University, Walton Hall, Milton Keynes, MK15 0BT
2: UK Centre for Astrobiology, University of Edinburgh, Edinburgh, EH9 3FD
3: SUPA, School of Physics and Astronomy, University of St Andrews, North Haugh, St Andrews KY16 9SS
4: Carl Sagan Institute, Department of Astronomy, Cornell University, Ithaca NY, 14853

**Direct Correspondence To:**
Duncan Forgan
dhf3@st-andrews.ac.uk

## Abstract

We address the possibility that intelligent civilisations that destroy themselves could present signatures observable by humanity. Placing limits on the number of self-destroyed civilisations in the Milky Way has strong implications for the final three terms in Drake's Equation, and would allow us to identify which classes of solution to Fermi's Paradox fit with the evidence (or lack thereof).

Using the Earth as an example, we consider a variety of scenarios in which humans could extinguish their own technological civilisation. Each scenario presents some form of observable signature that could be probed by astronomical campaigns to detect and characterise extrasolar planetary systems. Some observables are unlikely to be detected at interstellar distances, but some scenarios are likely to produce significant changes in atmospheric composition that could be detected serendipitously with next-generation telescopes. In some cases, the timing of the observation would prove crucial to detection, as the decay of signatures is rapid compared to humanity's communication lifetime. In others, the signatures persist on far longer timescales.

## 1 Introduction

The Search for Extraterrestrial Intelligence (SETI) owes a great deal of its observational and theoretical framework to the Drake Equation, which we quote in the form given by Walters et al (1980):

$$N = R_* f_g f_p n_e f_l f_i f_c L$$

Where $R_*$ is the mean star formation rate, $f_g$ is the fraction of stars that can host planetary systems, $f_p$ is the fraction of planetary systems that contain a habitable world, $n_e$ is the average number of habitable worlds per system, $f_l$ is the fraction of habitable worlds that contain life, $f_i$ is the fraction of inhabited worlds that contain intelligent civilisations, $f_c$ is the fraction of intelligent civilisations that choose to communicate, and $L$ is the typical civilisation lifetime.

Equally, SETI has been strongly influenced by Fermi's Paradox (Brin 1983, Webb 2002), which asks why humanity has no observational evidence for other intelligent civilisations, despite there being an apparent abundance of potential habitats for life and intelligence (Petigura et al 2013, Dressing & Charbonneau 2013) and ample time for civilisations to make their presence felt either via interstellar communication at lightspeed (Hair 2011) or exploration via physical probes at speeds comparable to those achieved by humanity's spacecraft (Nicholson and Forgan 2013).

Solutions to the Paradox typically require the product of the final three terms of the Drake Equation, $f_i f_c L$, to be small. This phenomenon is sometimes referred to as "the Great Filter", as it removes potential or existing civilisations from our view (Hanson 1998). As there are three terms to modify, there are three broad classes of solution to Fermi's Paradox, as elucidated in a review by Cirkovic (2009).

The first is dubbed the "Rare Earth" class, and suggests that $f_i$ is very small. While there may be many planets inhabited by single-celled or multicellular life $(f_l \sim 1)$, very few generate metazoan organisms that go on to found technological civilisations. The reasoning for this scenario is discussed in detail by Ward and Brownlee (2000) and more recently by Waltham (2015).

The second class requires us to consider how civilisations might limit their detectability, where $f_l$ and $f_i$ may be large, but $f_c$ is small. This may be due to agreements between existing civilisations to avoid the Earth (Ball 1973, Fogg 1987) or because the nature of reality requires there to be exactly one civilisation in the Universe, i.e. the Universe is a sophisticated simulation (e.g. Bostrom 2003). As this class challenges epistemology, it is difficult to consider scientifically. Also, solutions belonging to this class are often considered to be 'soft', as they require a uniformity of motive and behaviour that is difficult to cultivate over Galactic distances (Forgan 2011).

The third class demands that civilisations have short lifetimes ($L$ is small). Usually referred to as the 'Catastrophist' class, this requires civilisations to be extinguished either through natural means or through self-destruction. The Catastrophist class implies that civilisations are fragile, either due to external threats from devastating phenomena such as asteroid impacts, supernovae or gamma ray bursts, or that civilisations contain inherent social or structural flaws that prevent them from sustaining themselves over long time periods. If the destruction of civilisations is inevitable, then this will fundamentally limit the number of communicating civilisations present at any time, with obvious consequences for SETI (see e.g. Vukotic and Cirkovic 2008).

At the time of writing, all three classes of solution to Fermi's Paradox remain viable given our current lack of evidence. Current SETI searches rely on detecting intentional or unintentional signals at a variety of wavelengths (Reines and Marcy 2002, Howard et al 2004, Rampadarath et al 2011, Siemion et al 2013, Wright et al 2014). These searches generally set upper limits on the population and broadcast strength of communicating civilisations, but with only one civilisation in our sample (humanity), predicting which class of solution to Fermi's Paradox represents reality is extremely difficult.

If we cannot rely on the current data from SETI to constrain the last three terms of the Drake Equation and conclusively solve Fermi's Paradox, what other data can we turn to? Recent developments which constrain the earlier terms of the Drake Equation, such as advances in the detection and characterisation of extrasolar planets or exoplanets (Madhusudhan et al 2014) are likely to be crucial. Our improving ability to characterise potentially habitable worlds may begin to yield clues about intelligent agents and their (possibly deleterious) effect on planetary properties. Taking a pessimistic view of the changes we have made to the Earth's surface, atmosphere and its local environment, it seems possible that if extraterrestrial intelligences (ETIs) are common, observational evidence of intelligent self-destruction could also be common.

While it may be a morbid and depressing thought, looking for evidence of extraterrestrial civilisations that have undergone self-annihilation may be able to tell us much about the prevalence of intelligent life in the universe ($f_i$), as well as placing constraints on $L$. Indeed, this approach may present the best chance of finding any evidence of intelligent life beyond the Earth, as well as addressing two classes of solution to Fermi's Paradox.

The aim of this paper is to use the Earth as a test case in order to categorise the potential scenarios for complete civilisational destruction, quantify the observable signatures that these scenarios might leave behind, and determine whether these would be observable with current or near-future technology.

The variety of potential apocalyptic scenarios are essentially only limited in scope by imagination and in plausibility according to our current understanding of science. However, the scenarios considered here are limited to those that: are self inflicted (and therefore imply the development of intelligence and sufficient technology); technologically plausible (even if the technology does not currently exist); and that totally eliminate the (in the test case) human civilisation.

Only a few plausible scenarios fulfil these criteria:

i) complete nuclear, mutually-assured destruction
ii) a biological or chemical agent designed to kill either the human species, all animals, all eukaryotes, or all living things
iii) a technological disaster such as the "grey goo" scenario, or
iv) excessive pollution of the star, planet or interplanetary environment

Other scenarios, such as an extinction level impact event, dangerous stellar activity or ecological collapse could occur without the intervention of an intelligent species, and any signatures produced in these events would not imply intelligent life.

In the following sections, we will describe various ways that humanity may destroy its civilisation, and the observational signatures these events may produce. We will also discuss the timescales on which these signatures might persist, and prospects for their detection by present and future observations.

## 2 Destruction Channels for Intelligent Civilisations

There are many ways for humanity to end its civilisation. As we will see, the visibility and persistence of forensic evidence for a civilisation's death depends greatly on its cause. Our compilation of destruction channels is extensive, but not exhaustive.

### 2.1 Nuclear Annihilation

Perhaps the closest humanity has so far come to mutual annihilation is from the threat of global nuclear war (and the Doomsday Clock currently still stands at 5 minutes to midnight, according to the Bulletin of Atomic Scientists). While the tension of the Cold War may have passed, global nuclear arsenals are still large enough to destroy human civilisation.

Current estimates of nuclear weapons held around the world are of the order 6 million kilotonnes (kt) ($2.5 \times 10^{16}$ J), with the large majority of this total being in Russian hands, followed by the USA and then other nuclear nations at far lower totals (Kristensen and Norris 2014). Assuming a conflict that included the majority of these weapons, It is possible to make estimates of the effects of these weapons, on the assumption a global nuclear conflict occurs. Of course, the effects would not be evenly distributed, given that the majority of weapons would most likely be targeted at urban populations or military installations, but the global effects would be severe and widespread, even in places where there were no direct attacks.

Nuclear weapons produce a short, intense burst of gamma radiation with a characteristic double peak over several milliseconds (Weiss 2011) . These gamma flashes could be detected using the same techniques as for the detection of gamma ray bursts (GRBs, e.g. Kouveliotou et al 1993). In fact, the earliest detections of GRBs were initially thought to be nuclear weapons tests, due to their similarly short timescales and some similar spectral features.

However, GRBs are distributed evenly across the entire sky, pointing to origins beyond the Solar System, and indeed beyond the Milky Way. They are now thought to be powerful energetic events resulting from the mergers of compact objects such as neutron stars and black holes, or from the collapse of supermassive stars (see Fishman and Meegan 1995, Berger 2013 for reviews of the subject). Because of the extremely large energies released by GRB events (around $10^{44}$ J), these events are visible at extremely large distances. GRBs constitute the most distant objects ever observed by humanity. For example, the GRB 120923A has a measured redshift of z=8.5 (cf Tanvir 2013). Given that the world's nuclear arsenal is equivalent to around $10^{19}$ J of energy, the resulting radiation from its combined detonation would be much fainter than a typical GRB. If we assume that the energy is released on a similar timescale and with a similar spectrum to a GRB, a nuclear apocalypse is equivalent in bolometric flux to a GRB detonating around a trillion times closer than its typical distance. If we take a nearby GRB such as GRB 980425 (Galama et al 1998) which is thought to have detonated around 40 Mpc away, then we would expect a global nuclear detonation event to produce a similar amount of bolometric flux only 8 AU away!

Therefore, for us to be able to detect nuclear detonation outside the Solar system, the total energy of detonation must be at least nine orders of magnitude larger, i.e. the ETIs responsible for the event must engage in massive weapon proliferation and concurrent usage. However, the production of fallout from terrestrial size payloads, which persists for much longer timescales, may make itself visible in studies of extrasolar planet atmospheres.

For the purposes of estimating fallout, the weapon impacts are assumed to be evenly distributed across the entire land area of the planet ($1.5 \times 10^8$ km$^2$). This gives an equivalent of approximately one 25 kt ($10^{11}$ J) weapon per square kilometre of land area. This is of the same order of magnitude as the weapon used in the Semipalatinsk Nuclear Test (Imanaka et al. 2006), for which the effects of radioactive fallout were measured over time. However, given the local climatic conditions at this site (which were very windy) and the fact that our estimates include nuclear events every square kilometre, the effects are likely to be much worse than the results of this test, as reported by Sakaguchi et al (2006). From measurements of soil at a town near the test site and modelling of radionuclide decay chains, the dose rate due to fallout from the weapon test (not the dose from the blast itself) was shown to begin around $10^3$ microgray/hour, decaying to background levels after around 100 days.

Fallout products of fusion weapons are typically non-radioactive, though they do produce a low yield of energetic protons and electrons. Most fallout products from fission weapons are beta emitters and decay to other beta emitting isotopes (Hess 1964). Some radioisotopes produced by fission weapons are gamma emitters, but these have short half-lives. Ignoring the effects on the health of humans or other lifeforms (which would be severe), the deposition of a large amount of beta-radioactive material into the atmosphere would have a significant effect on atmospheric chemistry and would quickly ionise many atmospheric species, with high altitude nuclear tests increasing local electron density several times (Rothwell et al 1963). This would give ionised air the distinct blue or green of nitrogen and oxygen emission. Given that spacecraft and Earth based telescopes have detected (faint) nighttime airglow on Venus and Mars (Barth et al 1972, Krasnopolsky 1985) it may be possible to measure what would be considerably brighter airglow features in exoplanets, given that the order of magnitude increase in electron density caused by a nuclear war would generate an order of magnitude increase in airglow brightness (Meléndez-Alvira, et al. 1999). The brightest airglow feature in the visible spectrum on an Earth-like exoplanet would be the green oxygen line at 558 nm, which would be enhanced by global nuclear war to a photon flux of up to 1400 rayleighs (Greer, et al. 1986).

IR emission from exoplanets in their secondary eclipse phase has been measured by space-based telescopes (cf Charbonneau et al 2005, Baskin et al 2013) so in theory these measurements could be extended into the visible part of the spectrum in future, though this would require exquisite precision in our knowledge of the host star's properties, and would most likely be dominated by reflected light from the planet itself, especially in the blue-green spectral region. A ten-fold increase in brightness at 558 nm would potentially be observable with only a modest increase in sensitivity over instruments observing exoplanets today (Kreidberg et al. 2014), especially since the airglow maximum occurs well above the tropopause (Greer et al. 1986) and would therefore be observable above even a very cloudy planet. Airglow caused by fallout products would last for several years before decaying to unobservable levels.

The thermal effects of nuclear explosions also affect atmospheric chemistry. For every kilotonne in yield, approximately 5000 tonnes of nitrogen oxides are produced by the blast itself. Blasts from higher yield weapons will carry these nitrogen oxides high into the stratosphere, where they are able to react with and significantly deplete the ozone layer. Ozone can be detected in the ultraviolet transmission spectrum of an exoplanet, as can other oxygen molecules (Grenfell et al 2014, Misra et al 2014), and so the disruption of an exoplanetary ozone layer presents another potential observational signature.

Global nuclear war therefore potentially offers several spectral signatures that could be observed: a gamma flash, followed by UV/visible airglow and the depletion of ozone signatures. However, the aftermath of a global nuclear war will also act to obscure these spectral signatures. Ground-burst nuclear explosions generate a significant amount of dust that will be lofted into the atmosphere. Air-burst explosions do not generate dust, but still introduce particulates into the atmosphere. Atmospheric effects of nuclear warfare have been extensively modelled in climate simulations, the global consequences being known as "nuclear winter". Recent simulations have shown that even with reduced modern nuclear arsenals severe climate effects are felt for at least ten years after a global conflict, especially due to the long lifetime of aerosols lofted into the stratosphere (Robock et al. 2007). They show that the atmospheric optical depth is increased several times for several years. The worst effects are confined to the northern hemisphere given that the model includes conflict over the US and Russia, though the entire planet is affected to a lesser extent.

A nuclear winter would dramatically increase the opacity of the atmosphere. This process itself would be observable - if a planet observed with a previously transparent atmosphere (perhaps with an Earth-like spectroscopic signature) was observed again and the atmosphere was opaque, this would be a sign of a large dust event. However, such an event could also be caused by a large impact and therefore would not imply a civilisation had caused the disaster (though would be interesting in itself). If the atmosphere had not been observed before the event, it would simply seem like the planet had an extremely dusty atmosphere. What would be crucial is measuring the relative change in atmosphere as a result of nuclear detonation, hopefully with an added bonus of identifying a weak gamma ray or other high energy emission in the vicinity of the planet.

Hence, to confirm that a planet had been subject to a global nuclear catastrophe would require the observation of several independent signatures in short succession. One on its own is unlikely to be sufficient, and could easily be caused by any number of other processes on planets with potentially no biological activity whatsoever. There are cases beyond a global nuclear catastrophe that a space-faring civilisation might be able to inflict on itself, given that the destructive energy at their disposal would be far greater than nuclear weapons (Crawford and Baxter 2015), including redirecting asteroids. These would be far more destructive than nuclear warfare but would generate observable signatures different than those of a naturally occurring impact event.

## 2.2 Biological Warfare

Biological warfare involves the use of naturally occurring, or artificially modified, bacteria, viruses or other biological agents to intentionally cause illness or death. The use of a naturally occurring pathogen in a global conflict would probably have a limited net effect on a global population. The destruction would be self-limiting; once a population is reduced in size, transmission from host to host would become more difficult and the epidemic eventually ends. Artificially modified or created biological agents however, could potentially push a civilisation to extinction.

For example, small modifications to existing viruses could make them significantly more lethal. Jackson et al. (2001) describe how a simple modification to the mousepox virus, by introducing a gene that tells a mouse's immune system to shut down, caused it to kill all infected mice, including those vaccinated against the disease. It is not unfeasible to imagine a similar alteration to the human smallpox virus. The use of this in warfare would have devastating consequences, but would probably not cause complete human extinction as a result of the self-limiting nature of such an epidemic. Further modifications that allowed it to cross species barriers for example, could increase the magnitude of such a virus' effects, potentially causing a civilisation to completely destroy itself. Assuming a global conflict took place that made use of this method of warfare on a planet that hosts an intelligent civilisation, we pose the question of whether the self-destruction of that species, via this method, could be remotely observable.

If we assume that the time between the release of the engineered virus and its global spread is very short and that the virus is potent enough that a civilisation becomes fully extinct, the environmental impacts of this scenario can be assessed. The actions of anaerobic organisms cause biomass to decay, releasing methanethiol, $CH_3SH$ (via production of methionine) as one of the products. This can be spectrally inferred and has no abiotic source. For a population with a similar biomass to the present human population (currently, in terms of carbon biomass, ~$2.8 \times 10^{11}$ kg C - Walpole et al., 2012), the decay products can be estimated. Since the dry mass of a cell is approximately 50% carbon, the total human biomass would be $5.6 \times 10^{11}$ kg. With an estimated cell sulphur content of 0.3-1% (Pilcher, 2003), the maximum amount of S available to form $CH_3SH$ would be $5.6 \times 10^9$ kg. Following Pilcher (2003), if 10% of this S is incorporated into methionine, all of which is then converted into methanethiol, this would result in a total $CH_3SH$ flux of $\sim 10^8$ kg.

At the current biological production rate on Earth, this would be released to the atmosphere over a period of a year and would rapidly photodissociate, making this a very short-lived biosignature. One of the products of the decay of methanethiol is ethane ($C_2H_6$), which can be spectrally detected, but has an atmospheric lifetime under Earth-like conditions of < 1 year, leading to a short window of time for detection. Additionally, if carrion-eating species were unaffected, this would reduce the amount of organic matter available for microbial decay, further reducing the final biosignature.

However, if the engineered virus could cross species barriers, then the total amount of dead biomass could be as high as $6 \times 10^{13}$ kg (the total animal biomass on Earth - Groombridge & Jenkins, 2002), potentially producing $10^{11}$ kg of $CH_3SH$, which would enter the atmosphere over a period of ~30 years. It is likely that, due to its short atmospheric lifetime, this atmospheric $CH_3SH$ would still not produce a detectable signature. However, the associated $C_2H_6$ absorption signature between 11-13 μm may lend itself to remote detection. This signature would be deeper and therefore more readily detectable if the $CH_3SH$ production rate was higher (Domagal-Goldman et al., 2011).

Other decay products include $CH_4$, $H_4S$, $NH_3$ and $CO_2$. The most promising biosignature gas for global bioterrorism is $CH_4$. The $CH_4$ flux to the atmosphere is related to ethane production, potentially increasing the $C_2H_6$ absorption signature. In figure 1, changes in atmospheric $CH_4$ and $C_2H_6$ levels for two bioterrorism scenarios are plotted using an adapted form of a biosphere-atmosphere model from O'Malley-James et al. (2013). For the case where only humans can be infected, both signatures are short-lived, requiring observations to be taking place at exactly the right time for a detection to be made. In the case where the virus can cross species barriers leading the the total annihilation of animal life, persistently high levels of these gases could make a detection more likely.

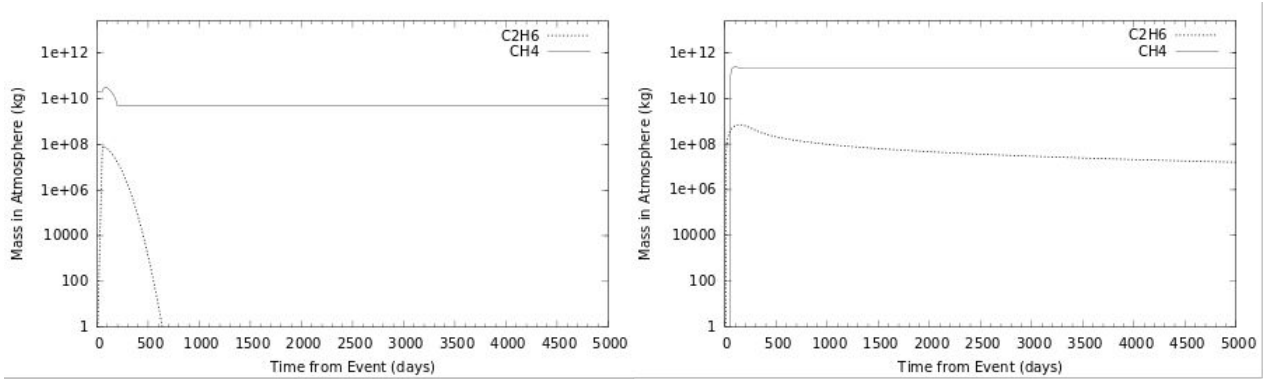

*Figure 1. (a) For the case where an engineered epidemic causes full extinction of the human population, C$_2$H$_6$ and CH$_4$ biosignature changes are short-lived and would need to be observed within 1 year of the event. (b) An epidemic that rapidly kills all animals could potentially be detectable years after the event.*

**2.3 Destruction via 'Grey Goo'**

The terrestrial biosphere offers many examples of naturally occurring nanoscale machines. Feynman (1960) extolled the advantages of engineering at atomic scales. In *Engines of Creation*, Drexler (1986) described "nanotechnology" as a means of fabricating structures at nanoscales using chemical machinery. While the word now has a broader meaning, we can still consider the possibility that such a machine can be sufficiently general-purpose to be able to make a copy of itself.

Following Phoenix and Drexler (2004) we define an engineered system that can duplicate itself exactly in a resource-limited environment as a *self-replicator*. (NB: This strict definition excludes biological replicators, as they are not engineered). The engineers of such machines have two broad choices as to what resources the self-replicator might use: resources that are naturally occurring in the biosphere, and resources that are not. Engineers that make the former choice run the risk of a "grey goo" scenario, where uncontrolled self-replication converts a large fraction of available biomass into self-replicators, collapsing the biosphere and destroying life on a world. This may be an accident or failure of oversight, or it may be due to a deliberate attack, where the replicators are specifically designed to destroy biomass (what Freitas 2000 refers to as "goodbots" and "badbots" respectively). In *Engines of Creation,* Drexler (1986) notes:

*"Replicators can be more potent than nuclear weapons: to devastate Earth with bombs would require masses of exotic hardware and rare isotopes, but to destroy all life with replicators would require only a single speck made of ordinary elements. Replicators give nuclear war some company as a potential cause of extinction, giving a broader context to extinction as a moral concern."*

Freitas (2000) places some important limitations on the ability of replicators to convert the biosphere into "grey goo" (land based replicators), "grey lichen" (chemolithotrophic replicators), "grey plankton" (ocean-borne replicators) and "grey dust" (airborne replicators). With conservative estimates based on contemporary technology, it is suggested that if the replicators are carbon-rich, around a quarter of the Earth's biomass could be converted as quickly as a few weeks. Equally, Freitas (2000) estimates the energy dissipated by carbon conversion, implying that subsequent thermal signatures (local heating and local changes to atmospheric opacity) would be sufficient to trigger local defence systems to combat gooification. For example, In the case of malevolent airborne replicators, a possible defensive strategy is the deployment of non-self replicating "goodbots" which unfurl a dragnet to remove them from the atmosphere.

Phoenix and Drexler (2004) emphasise that all these variants of the grey goo scenario are easily avoidable, provided that engineers design wisely (and that military powers exercise restraint). Indeed, they indicate that fully autonomous self-replicating units are not likely to be the most efficient design choice for manufacturing, and that having a central control computer guiding production is likely to be safer and more cost-effective. Provided that the control computer is not separated by distances large enough to introduce time-lag, as would be the case on interplanetary scales, this seems to be reasonable.

However, this still leaves the risk of replicator technology being weaponised. We will assume, as we do throughout this paper, that prudence is not a universal trait in galactic civilisations, and that grey goo is a potential death channel that might be detected.

So what signatures might a grey goo scenario produce? If a quarter of the Earth's biomass is converted into micron sized objects, how would this affect spectra of Earthlike planets? This situation shares several parallels with the nuclear winter scenario described previously. In the case of grey goo, we may expect there to be a substantially larger amount of "dust", as well as a fixed grain size. This will be deposited as sand dunes or suspended in the atmosphere, with similar spectral signatures as previously discussed.

Depending on the grain size of the dunes, it may be possible to observe a brightness increase as the angle measured by the observer between the illumination source (the host star) and the planet decreases towards zero on the approach to secondary eclipse.

Surfaces that are composed of a large number of relatively small elements packed together will produce significant shadowing. This shadowing increases as the angle between the surface and an illumination source increases. As the angle decreases towards zero, these shadows disappear, resulting in a net increase in brightness. This is sometimes described as the opposition surge effect, or the Seeliger effect in deference to Hugo von Seeliger, who first described it. Seeliger saw this shadow-hiding mechanism in Saturn's rings, which grow brighter at opposition relative to the planetary disc (see e.g. Fischer et al 2011 for recent measurements which show this). Coherent backscattering of light also plays a role in this brightening effect.

This phenomenon is observed in the lunar regolith (Buratti et al 1996), so it seems reasonable to expect that this phenomenon would also act in artificially generated regoliths such as those we might expect from a grey goo incident. During exoplanet transits, it may be possible to detect an increase in the brightness of the system as the planet enters secondary eclipse. The Moon's brightness increases by around 40% as it moves towards the peak of opposition surge, so it may well be the case that grey-goo planets produce opposition surges of similar magnitude. Buratti et al (1996) note that the wavelength dependence of the surge is relatively weak, which would suggest that near-IR observations may be sufficient to observe this phenomenon.

On what timescale might we expect this artificial nano-sand to persist on a planetary surface? If the planet has an active hydrological cycle, airborne replicators will be incorporated into precipitation and delivered to the planet's surface. The Earth's Sahara desert transports away of order a billion tonnes of sand per year (Goudie and Middleton 2001). Deposition into rivers and streams may deliver the material to oceans, and eventually the seabed, effectively removing it from view at interstellar distances. This material will be subducted into the mantle and reprocessed on geological timescales, removing all trace of engineering. Using Freitas (2000)'s estimate of available biomass, and assuming the nano-sand can be processed out of view at a few billion tonnes per year (which we propose as an upper limit) then this suggests that a goo-ified planet may require several thousand years to refresh its surface. It is likely that processing rates may be accelerated or impeded by other physical processes, but it seems to be the case that goo-ified planets remain characterisable as such over timescales comparable to that of recorded human history.

**2.4 Pollution**

**2.4.1 Pollution of the Host Star**

It is clear, from observations of our Solar system and exoplanet systems, that the planet formation process will lead to natural "pollution" of host stars, whereby "pollution" we mean accretion of planets and planetary debris (see e.g. Li et al 2008).

However, it is also possible that civilisations may wish to deposit waste material on their host star, especially if such waste is deemed hazardous. The byproducts of nuclear fission are certainly hazardous, and remain so on significantly long timescales - proposed terrestrial storage systems for high-level waste products are designed with a view to being secure for at least 10,000 years (see e.g Kerr 1999). As a star's gravitational well will likely be the deepest in the system, it is also energetically efficient to launch material on a stellar-intercept trajectory. If the outer convective layer of the star is not too deep, then it seems likely that this waste will remain near the stellar photosphere, and hence be detectable in stellar spectra. Shklovskii & Sagan (1966) went as far as suggesting that civilisations might "salt" their star by deliberately placing rare isotopes in the stellar photosphere to act as an interstellar marker for other civilisations to detect.

Whitmire & Wright (1980) considered civilisations carrying out neutron fission of plutonium-239 and uranium-233 on industrial scales, and compared the elemental abundances of their byproducts to those of the Sun. They found that in both reactions, praseodymium (atomic number 59) a relatively under-abundant element in the Sun, becomes artificially enhanced through the addition of nuclear waste. However, they note that the outer convective layer of solar-type stars is probably too deep to maintain a strong signal, as it is likely most of the pollutants will be transported far from the photosphere. Stars with masses greater than around 1.5 solar masses will have thin surface convection zones, and as such should be able to present strong signals, even with a relatively small amount of pollutant (equivalent to exploiting around 0.5% of the Earth's uranium-238 for plutonium-239 fission).

These signals do not distinguish between living and extinct civilisations.  At best, we can consider the extreme case, where large amounts of artificial pollution begin to alter the star's equilibrium structure.  Cody and Sasselov (2004) produced stellar structure models which include the effect of pollution in the surface convective zone.  Their modelling considered pollution as a general increase in the metallicity of the zone (NB: astronomers define metallicity as the abundance of elements that are not hydrogen or helium).  In general, they find that the convective zone deepens as pollution increases.  However, the distance between the stellar core and inner boundary of the zone does not decrease - concordantly, the radius of the star increases, allowing the effective temperature of the photosphere to drop while maintaining a roughly constant luminosity.

This would immediately suggest compiling a sample of stars that are bright, cool and slightly larger than expected as an initial step to search for this particular death channel.  However, the mass of pollutants investigated by Cody and Sasselov in their analysis is typically of the order of 10 Earth masses! Such a vast quantity of pollution is likely to have its origins in the planet formation phase, rather than through intelligent intervention.  It is therefore necessary to obtain abundance data for this putative sample to identify unnaturally abundant elements such as the fission byproducts discussed above.

It is extremely unclear how pollution of the host star might affect habitability of any planets in orbit.  From observations of comets such as ISON making close approaches to the Sun (Ferrin 2014) there does not appear to be any corresponding increase in stellar activity or coronal mass ejections (CMEs), but obviously a comet and a container of nuclear waste have extremely different properties.  However, if the effective temperature of the star changes as a result of the deepening surface convection zone, while the luminosity remains constant, the habitable zone boundaries of the system will move inward (Kopparapu et al 2013)[1], and consequently planets that were previously habitable may become too cold to support life.  Such an event would require drastic levels of pollution to achieve.

---

[1] http://depts.washington.edu/naivpl/sites/default/files/index.shtml, accessed 03/07/15

So, is stellar pollution a useful marker for exopocalypse? It seems to be the case that stellar pollution is an indicator of the presence of intelligence, but as has been mentioned, it does not clearly distinguish between living and extinct civilisations. Stars that have extremely large abundances of fission byproducts, along with planets near to, but beyond the outer edge of the local habitable zone, may be good candidates for sites of dead civilisations, but such evidence is hardly conclusive.

**2.4.2 Pollution of the Host Planet**

Detecting artificial pollutants in exoplanet atmospheres is not a new idea in SETI (Schneider et al 2010). In particular, the presence of CFCs in the Earth's atmosphere is a telling indication of industry in the Earth. It has been demonstrated that the James Webb Space Telescope could detect CFCs at ten times the Earth's levels in the atmospheres of Earthlike exoplanets (Lin et al 2014).

With signal lifetimes ranging between 10 and $10^5$ years, it is possible that a heavily CFC-polluted exoplanet could remain detectable for a substantially long time. Also, the detection of multiple CFC signatures could provide a form of chemical dating. Short-lived CFCs indicate active production, and hence an active civilisation, whereas the absence of short-lived CFCs alongside strong signals of long-lived CFCs would be a good indicator of either a) an extinct industrial civilisation, or b) a surviving industrial civilisation that has ceased producing CFCs.[2]

A third possibility is that civilisations have surpassed (or succumbed to) a technological Singularity (Kurzweil 1999). This is less the death of a civilisation than its transmutation into something fundamentally different. One might assume that post-Singularity civilisations cease environmental pollution due to exponentially increasing technological innovation, but as always assumptions regarding the behaviour of postbiological civilisations must be made carefully (cf Dick 2003).

As CFC pollution is covered in detail in other work, especially Lin et al (2014), we will not repeat it here, but this technomarker is likely to be one of the most straightforward means of detecting extraterrestrial intelligence, and one of the most straightforward measurements of civilisation prudence and forward planning.

---

[2]

**2.4.3 Pollution of the Planetary Orbital Environs**

Humanity has been adding artificial material to Earth's orbit since the launch of Sputnik in 1957 by the Soviet Union. Current activities in Earth orbit - placing artificial satellites, constructing space stations, launching probes to other planets and beyond - typically result in the production of debris, ranging in size from paint flecks to spent engine stages. NASA's Orbital Debris Program Office currently tracks a total of 21,000 objects greater than 10cm in size in low Earth orbit (LEO) classifiable as debris, with estimates of populations below 10cm substantially higher[3].

Many orbits after launch, spacecraft still present a danger of further debris production, either through spacecraft-spacecraft collisions, collisions with other debris, or deliberate action, such as the intentional destruction of the Fengyun 1C satellite by the People's Republic of China, which created over two thousand additional debris fragments (Johnson et al 2008). If the debris produced by a collision remains above a critical size (of order one metre), this further escalates the risk of future collisions. It is possible this may set up a collisional cascade where one collision generates a chain reaction of collisions, with potentially devastating consequences. In an extreme scenario, it is possible that LEO may become impassable to spacecraft, which would have severe consequences for human civilisation.

Kessler and Cour-Palais (1978) considered the then-population of satellites in orbits up to altitudes of 4000 km. For any given spacecraft, the impact rate depends on the local density of objects, the mean relative velocity, and the mean cross-sectional area of the objects. Using the then-exponential growth of catalogued objects in orbit, and extrapolating to obtain future populations, Kessler and Cour-Palais (1978) predicted that by the year 2000 the local density of spacecraft above a critical size would be sufficient to generate a cascade. This debris would eventually form a ring system around the Earth, until atmospheric drag removes the material from orbit. This collisional cascade is now commonly referred to as *Kessler syndrome*. While the onset of Kessler syndrome by the year 2000 did not come to pass, it may still occur in the future.

The current LEO environment is dominated by explosion fragments, but recent modelling indicates that over the next century, the debris produced by collisions will dominate the field, and without mitigation of this debris, will present a serious risk to human activity, especially at LEO (Liou 2006).

---

[3] http://orbitaldebris.jsc.nasa.gov

It is conceivable that a similar fate may befall civilisations at our stage of development. The evasion of space debris by say, the International Space Station (ISS), requires significant course correction[4]. If a civilisation has failed, then the ability of large space-based structures to avoid debris strikes and meteoroid hits will be diminished, and a Kessler syndrome scenario could be achieved with a lower density of objects. Even if a civilisation does not fail, malicious activity could result in devastation of the low orbit environment, which in itself could precipitate collapse.

Could such a disaster be viewed from interstellar distances? It seems to be the case that the debris would form a ring system, which could be detectable from transits depending on the ring's surface density (Barnes and Fortney 2004). Given the ring optical depth ($\tau$) and the sine of the ring tilt to our line of sight ($\beta$), the rings will create a dip in the transit curve proportional to $e^{-\tau/\beta}$ as a result of extinction. Depending on the grain size of the debris, we may expect scattering to reduce the transit curve even further. Equally, diffraction can reduce the transit depth for relatively large grains (whereas small grains diffract isotropically). As the debris will not be as icy as the ring systems we observe in the Solar System, it is possible that the reflective properties of the debris may present unusual polarisation signatures.

Detectability of rings via transits appears to be independent of distance from the star, although most calculations assume the rings encircle a giant planet. These appear to be within the reach of Kepler observations, although true Saturn analogues are likely to be infrequent in the Kepler sample (Barnes and Fortney 2004). If we expect a debris ring to extend to (at most) 2 Earth Radii from a terrestrial planet, with an optical depth of less than unity, then it seems that current observations are not capable of this task.
However, another promising avenue could arise from asterodensity profiling (Zuluaga et al 2015). This technique attempts to measure the stellar density by using measuring the ratio of the planet semi major axis to the stellar radius. This ratio is related to the stellar density through Kepler's 3rd Law, and is determined from the transit curve. If orthogonal measurements of the stellar density are made using e.g. asteroseismology data, the presence of planetary rings produces a discrepancy between the two measured stellar densities (what is known as the *photo-ring effect*).

---

[4] http://www.nasa.gov/mission_pages/station/news/orbital_debris.html , accessed Oct 2014

The recent detection of substantial ring features in the transit curve of J1407b are an encouraging step forward (Mamajek et al 2015). While the nature of the rings' host remains unclear (it may be a giant planet, brown dwarf or low mass star), it demonstrates that circumplanetary ring and disk systems are in principle detectable, and that increased sensitivity and cadence will allow similar observations of lower mass systems in future.

Even with future instrumentation and substantial amounts of debris, ring systems will need to be either frequent or long-lived to be detectable.  Unless the debris system enters an equilibrium state, where there is a sufficient supply of large objects to feed the production of smaller objects, the ring system will dissipate as the particle's orbits are decayed by Poynting-Robertson drag, as well as drag originating from the planet's exosphere (Gaudi et al 2003). This can occur on timescales as short as 100,000 years, and is potentially even shorter for terrestrial planets.

Is this signature unequivocally from dead civilisations? Earth is an example of a living civilisation with space debris, albeit not yet formed into a ring system. Current plans to mitigate the threat of Kessler syndrome require spacecraft to de-orbit potential impactors, either through direct capture, such as the Swiss Space Agency's CleanSpace One satellite (Richard et al 2013) which will collect debris and then de-orbit with the destruction of both satellite and debris, or through long-range orbit modification, such as Mason et al (2011)'s proposal to use laser pulses to induce extra drag on debris. It is likely that if Earth evades Kessler syndrome, some debris will remain, although at levels potentially undetectable using our current instrumentation. At best, we can say that detection of an artificial ring system around a habitable planet implies a civilisation has undergone a Kessler syndrome event - whether such an event is catastrophic will rely on the density of debris, and the civilisation's dependence on the orbital environment for sustaining itself. It could be argued that civilisation might continue in the absence of satellite technology or general access to space, albeit at a significant disadvantage.

It is equally possible that habitable planets without intelligent civilisations may possess natural rings. Simulations of the latter stages of planet formation indicate that giant impacts, such as that which created the Earth's moon, are common (cf Jacobson and Morbidelli 2014) and hence it is possible that impacts in differing configurations may create debris fields around terrestrial planets that form rings.  Discrimination between natural and artificial rings will require significant modelling of the expected grain size distribution of both types, as well as the material properties of the grains, as these will affect both photometry and polarimetry measurements.

**2.5 Total Planetary Destruction**

Finally, it is not inconceivable that a civilisation capable of harnessing large amounts of energy could unbind a large fraction (or all) of a planet's mass.  Kardashev Type II civilisations (Kardashev 1964) wishing to build a Dyson sphere require this capability to generate raw materials for the sphere - it is estimated that to create a Dyson sphere in the Solar System with radius 1 AU would require the destruction of Mercury and Venus to supply sufficient raw materials (Dyson 1960, Vadescu and Cathcart 2000).

Equally, civilisations with access to this level of energy control and manipulation may decide to use it maliciously, destroying large parts of a planetary habitat while it is still occupied, and in the extreme case destroying the planet completely.  This would release a significant fraction of the planet's gravitational binding energy.

The Earth's binding energy is of order $10^{39}$ ergs.  This is again several orders of magnitude fainter than a typical supernova or GRB of $10^{51}$ ergs, but is strong compared to the solar luminosity - the Sun would require several days to radiate the same quantity of energy.  This would likely produce a gamma ray signature even stronger than expected from the nuclear winter scenario described previously, and we may expect afterglows similar to those observed in other astrophysical explosions.

The destruction of an orbiting body will produce a ring of debris around the central star, in a manner analogous to the production of rings when solid bodies cross the Roche limit of a larger body.

The subsequent evolution of this material will be similar to that of the debris discs.  The remnants of the planet formation process, debris discs have been detected around a variety of stars, and the behaviour of grains of differing sizes under gravity and radiation pressure has been modelled in detail (see e.g. Krivov et al 2013).

It is likely that, if a terrestrial planet has been destroyed, the debris will be principally composed of silicates, and as such any detection of refined or engineered materials is unlikely, even if such matter survives the planet's demise untouched.

The fate of the material depends largely on the local gravitational potential and the local radiation field, as well as the grain size distribution of the debris. Grains below the "blow-out" size - typically a few microns - will be removed from the system via radiation pressure. Neighbouring planets may collect some of the remaining debris in resonances (cf Mustill and Wyatt 2011) while the debris grinds into material of sufficient grain size that it either loses angular momentum through Poynting-Robertson drag and is consumed by the central star or a neighbouring planet, or gains momentum through radiative forces and is removed from the system.

In any case, this death channel does not appear to be amenable to detection by Earth astronomers. If we are fortunate to witness the instant of destruction, then we may be able to speculate on the energies released in the event, and search for a natural progenitor of such energy, i.e. another celestial body. Giant impacts between planet-sized bodies will produce the required energies to unbind or destroy one of the objects, as was the case for the impact which formed the Earth-Moon system (cf Canup 2008). If such efforts fail, and no other explanation fits the observations, then we may tentatively consider extraterrestrial foul play.

The timescale for observing destruction as it happens will be short - perhaps a few days. The debris can be expected to persist for several centuries, but observing this is unlikely to elucidate its origins as a destroyed planet.

## 3 Prospects for Observing Civilisation Destruction

| Death Channel | Detection Method | Signature of Active Civilisation | Signature of Dead Civilisation | Detection Timescale (yr) |
|---|---|---|---|---|
|  |  |  |  |  |

| Nuclear Detonation | Gamma ray detection, Transit spectroscopy | Y | Y | 0-5 years |
|---|---|---|---|---|
| Bioterrorism | Transit spectroscopy | Y | Y | 1-30 years |
| Grey Goo | Transit spectroscopy and photometry | N | Y | >1,000 years |
| Stellar Pollution | Asteroseismology, stellar abundance studies | Y | Y | >100,000 years (depending on stellar convection) |
| Planetary Pollution | Transit spectroscopy (IR) | Y | Y | 10-100,000 years |
| Orbital Pollution (Kessler Syndrome) | Transit spectroscopy and photometry | Y | Y | <100,000 years |
| Total Planetary Destruction | Debris Disk Imaging (IR) | Y | Y | <100,000 years |

**Table 1**: A summary of the destruction channels discussed in this paper

As we have seen, the observational signatures of self destroyed civilisations decay on a variety of timescales (Table 1). Some decay so rapidly it is unlikely that we will observe them without a large amount of serendipitous measurements.

We also rely heavily on the characterisation of exoplanet atmospheres, which is not without its difficulties. In particular, we require detailed characterisation of Earthlike planets around main sequence stars, a feat that is unlikely to be achieved until the launch of the James Webb Space Telescope . Even then, JWST is more likely to achieve this around low mass M stars at distances of order 10 parsecs (Batalha et al 2014, Barstow et al 2015). On timescales of a few decades, the TESS (Ricker et al 2014) and PLATO (Rauer et al 2014) exoplanet transit surveys, along with the upcoming Extremely Large Telescopes (ELTs, cf

Udry et al 2014), will produce a step change in the quantity and quality of target planets, as well as the sensitivity and resolution of the spectroscopic data.

Further constraints derive from the ability to model exoplanet atmospheres in the absence of the technological signatures being discussed. Retrieval of atmospheric composition from current datasets is non-trivial, and the uncertainty of measurements depends on the retrieval algorithm and its assumptions. As such, the validity of some inferences is a matter of debate (Burrows 2014). Any atmospheric death signature will fall prey to the same issues.

Detection techniques that focus on transit photometry rather than transit spectroscopy are likely to be more effective in the near term. The appearance of artificial ring systems as a result of civilisation destruction appears to be well suited to future space missions. However, more sophisticated models of ring systems may be required to achieve this goal.

On a positive note, all the above detection methods are suited to "piggyback" operations, or mining of freely available archive data (cf Wright et al 2014). They are complementary to other forms of SETI search, and rely on astronomical data that is already of intense scientific interest. It seems clear that as instrumentation improves, SETI scientists will be able to take advantage and produce good constraints on the number of self destroyed civilisations in the Solar Neighbourhood.

**4 Summary**

We have outlined several methods for detecting extraterrestrial intelligences that have destroyed themselves. The probability of detection depends sensitively on the means by which a civilisation suffers annihilation. In most cases, the destruction of a technological civilisation leaves atmospheric traces that persist for a short time, requiring observations to be relatively serendipitous.

Civilisations which initiate a nuclear catastrophe produce strong but relatively brief signatures of their destruction, which are partially masked by the dust thrown into the atmosphere by multiple nuclear detonations. Victims of bioterrorism produce powerful atmospheric signatures of decaying organic matter, but these dissipate on timescales of a few decades.

Nanotechnology, if left to consume large quantities of biomass, will produce extremely dusty atmospheres and large amounts of surface "desert" which could yield detection via shadow-hiding during transit photometry observations.  These signatures would remain on thousand year timescales, and potentially far longer depending on the rate that the nano-sand is delivered to the ocean floor.

Pollution of the host star leaves distinct traces of radioactive elements in the star's photosphere.  Depending on the depth of the star's convective layer, the signals could remain on thousand year timescales.  Excessive pollution may deepen the convective layer and reduce the effective temperature of the star, leaving once habitable planets stranded beyond the outer habitable zone boundary.

Pollution of the planet's atmosphere persists on a variety of timescales depending on planetary properties, from decades to millennia.  Pollution of the planet's local orbital environment produces rings of debris that may last for a few thousand years before being deorbited by atmospheric drag, but would be challenging to detect via transit photometry.

In the most extreme case, the destruction of planetary bodies produces a very strong initial radiation burst from the unbinding of the planet, followed by the production of debris disc rings that may indicate signs of artificial construction in their chemical composition.

In closing, it is clear that some observational signatures of self-destructive civilisations are currently amenable to astrophysical observations, but these will be challenging, and in some cases will require a degree of luck in observing at the correct time.  However, these detection techniques are relatively cheap, as they dovetail neatly with current astronomical surveys.  In time, the first evidence of extraterrestrial intelligence may come to us from the remains of less prudent civilisations.  In doing so, such information will bring us not only knowledge, but wisdom.

**5 Acknowledgements**

The authors would like to thank Ian Crawford for his insightful and fair review of this manuscript.  This paper was birthed during the "Building Habitable Worlds" workshop, hosted by the UK Centre for Astrobiology at the University of Edinburgh, and funded by the Scottish University Physics Alliance (SUPA).  DF gratefully acknowledges support from the ECOGAL ERC advanced grant, and the STFC grant ST/J001422/1.

arXiv:1412.1048